\documentclass[a4paper,twocolumn,english,pra,superscriptaddress]{revtex4}
\usepackage[T1]{fontenc} \usepackage[latin1]{inputenc}

\makeatletter

\usepackage{graphicx} \usepackage{amsmath} \usepackage{amssymb}
\usepackage{babel} \makeatother

\begin{document}

\title{Casimir forces between defects in one-dimensional quantum
liquids}

\author{A.~Recati} \affiliation{Institute for Theoretical Physics,
Universit\"at Innsbruck, Technikerstrasse 25, A-6020 Innsbruck,
Austria.}  \affiliation{CRS BEC-INFM, Povo and ECT$^\star$,
Villazzano, I-38050 Trento, Italy.}

\author{J. N. Fuchs} \affiliation{Laboratoire de Physique des Solides,
Universit\'e Paris-Sud, B\^atiment 510, F-91405 Orsay, France.}

\author{C. S. Pe\c ca} \affiliation{Institute for Theoretical Physics,
Universit\"at Innsbruck, Technikerstrasse 25, A-6020 Innsbruck,
Austria.}

\author{W. Zwerger} \affiliation{Technische Universit\"at M\"unchen,
James Franck Strasse, D-85748 Garching, Germany.}

\begin{abstract}
We discuss the effective interactions between two localized
perturbations in one-dimensional (1D) quantum liquids. For
non-interacting fermions, the interactions exhibit Friedel
oscillations, giving rise to a RKKY-type interaction familiar from
impurity spins in metals. In the interacting case, at low energies, a
Luttinger liquid description applies. In the case of repulsive
fermions, the Friedel oscillations of the interacting system are
replaced, at long distances, by a universal Casimir-type interaction
which depends only on the sound velocity and decays inversely with the
separation. The Casimir-type interaction between localized
perturbations embedded in a fermionic environment gives rise to a long
range coupling between quantum dots in ultracold Fermi gases, opening
a novel alternative to couple qubits with neutral atoms.  We also
briefly discuss the case of bosonic quantum liquids in which the
interaction between weak impurities turns out to be short ranged,
decaying exponentially on the scale of the healing length.
\end{abstract}

\date{\today}

\maketitle
\section{Introduction}
Interactions between localized defects which are mediated by the
continuum they are embedded in, play an important role in many areas
of physics.  Typical examples are the RKKY interaction between spins
in a Fermi liquid or the interaction between vortices in
superfluids. In the present work, we discuss interactions between
impurities in 1D quantum liquids.  This study is motivated by the
recent realization of strongly interacting ``atomic quantum wires''
with ultracold gases of both bosonic \cite{Paredes,Weiss,Moritz} and
fermionic atoms \cite{Esslinger} and the proposal \cite{Recati}, that
single atoms in optical traps which are embedded in a superfluid
reservoir allow to realize an atomic analog of a quantum dot with a
tunable coupling to the environment.  Such quantum dots may be used to
store qubits, which, under certain conditions, can be completely
decoupled from their environment.  Arrays of these dots thus appear as
ideal candidates for quantum information processing. It is therefore
of considerable interest to study the induced interactions of such
dots, mediated by the environment they are embedded in. Similar
questions arise also for quantum dots in solid state realizations,
e.g. in carbon nanotubes \cite{Marcus}, where the interaction is
mediated by electrons in the intervening wire.

Quite generally, for both bosons and fermions, the low energy
properties of a gapless 1D quantum liquid are described by the so
called Luttinger liquid (LL) phenomenology \cite{Haldane1,Haldane2}:
the effective theory is a hydrodynamic energy functional characterized
by the velocity of sound $u$ and the Luttinger interaction parameter
$K$.  In particular, for fermions, $K=1$ corresponds to the
non-interacting case, while $K<1$ for repulsion. For repulsive bosons,
in turn, one has $K>1$, with $K\to\infty$ in the limit of weak
interactions, where a Gross-Pitaevskii or Bogoliubov approximation
applies. As shown by Kane and Fisher \cite{KaneFisher} 
(see also \cite{Enss} for a recent discussion), the
interaction of a single impurity with a LL depends crucially on the
value of $K$: for $K>1$, the impurity is irrelevant for the low energy
properties. A 1D Bose liquid is therefore effectively superfluid,
although there is no true condensate \cite{Popov,KaneFisher}. For
$K<1$ the impurity changes the ground state of the liquid in a
non-perturbative way, effectively cutting it into two disconnected
parts. In this case, we will see that the induced interaction between
two impurities is essentially a Casimir-like effect. Indeed, at low
energies, two impurities at distance $r$ define a box with reflecting
boundary conditions for the phonon modes of the quantum liquid, which
leads to an attractive Casimir interaction energy proportional to
$u/r$ with $u$ the sound velocity.  The case $K=1$ is marginal and
corresponds to a non-interacting Fermi gas in 1D or - equivalently -,
a system of hard core bosons, the Tonks-Girardeau gas \cite{Girardeau,
Weiss, Paredes}.  In the following we will study the interactions
mediated by the 1D quantum liquid between two impurities for the
various cases, including fermions with spin. We focus our analysis on
the case of {\it static} impurities, while the situation of a slow
time dependence, relevant for atomic quantum dots, where the
interactions depend on the internal states is only discussed
qualitatively at the end of the paper.

\section{One-dimensional fermionic liquid}
\subsection{Non-interacting fermions}\label{nonint}
Before considering the generic situation of impurities embedded in a
sea of interacting particles, we first address the marginal case $K=1$
of non-interacting fermions.  For simplicity we start from a gas of
$N$ non-interacting spinless fermions in the presence of two localized
impurities separated by a distance $r$.  Considering cold gases in
atomic quantum wires, the solution of this problem is not just an
academic exercise. Indeed, since fermions in a single hyperfine state
have no s-wave interactions due to the Pauli principle, they realize
an ideal Fermi gas at sufficiently low temperatures. We assume that
the particles are contained in a periodic box of length $L$ with
average density $\rho_0\equiv N/L$. The (grand canonical) partition
function of the liquid at a given temperature $T$ may be expressed in
terms of a functional integral over the Grassman fields
$(\bar{\psi},\psi)$ representing the fermions:
\begin{equation} 
Z=\int \mathrm{D} \bar{\psi} \mathrm{D}\psi
  \exp(-S_\mathrm{FL}-S_{i}).
\label{eq:statsum}
\end{equation}
The corresponding action of an ideal gas is
\begin{equation}
S_\mathrm{FL}=\int_{0}^{\beta} \!\!\mathrm{d} x\,\mathrm{d}\tau
  \left[\bar{\psi}\,\partial_{\tau}\psi-\left(\frac{1}{2m} \nabla
  \bar{\psi} \nabla\psi -\mu\,\bar{\psi}\psi \right)\right],
\label{eq:SFL}
\end{equation} 
where $\tau$ is the imaginary time running from $0$ to $1/T=\beta$,
and $\mu$ is the chemical potential (we use units such that
$\hbar=k_B=1$).  The fields are anti-periodic in imaginary time. For
short range interactions, appropriate for cold atoms, the interaction
of the impurities with the liquid can be described by an additional
contribution
\begin{equation}
S_{i}=\int_{0}^{\beta}\mathrm{d}\tau\!\sum_{\alpha=1,2}g_{\alpha}\bar{\psi}
(x_{\alpha})\psi(x_{\alpha}),
\label{eq:Sint}
\end{equation} 
proportional to the local density at the impurity positions
$x_{\alpha}$. Here, the index $\alpha=1,2$ labels the impurities,
while the coupling constants $g_{\alpha}$ describe the strength of
collisions between the atoms in the liquid and the impurities. The
expression for $S_{i}$ is based on assuming an effective
pseudopotential for the interaction between the impurity and the
quantum liquid.  More precisely, the interaction should be replaced by
a spatial integral of the detailed impurity potential with the
microscopic density operator of the liquid. 
%In what follows, we
%consider both the actions (\ref{eq:SFL}) and (\ref{eq:Sint}) as the
%effective low energy expressions with the fields $(\bar{\psi},\psi)$
%containing the low energy part of the particle fields limited by the
%cutoff frequency $\omega_{c}$. 
In the present section dealing with non-interacting fermions, there is no need of 
a high energy cutoff, as one can directly work with the well behaved microscopic 
theory. However, in order to discuss the low energy behavior and to make contact 
with the following sections dealing with interacting fermions using the Luttinger 
liquid phenomenology, we introduce a high energy cutoff $\omega_c$. Its value can be
estimated as $\omega_{c}\sim \textrm{Min}\{u/l_{0},\mu\}$, where $u$
is the characteristic velocity of excitations ($u=v_F$ in an ideal Fermi gas), 
$l_0$ is the impurity size and $\mu=p_{F}^2/2m$ is the chemical potential (at zero 
temperature) with $p_{F}=mv_F \equiv \pi \rho_0$, the Fermi momentum. 
As will become clear from our results
below, the coupling constants $g_{\alpha}$ are then - up to a factor
$v_{F}$ - identical with the dimensionless backscattering amplitudes
$f_{1,2}(\pi)$ for fermions at the impurities. Microscopically, they
are thus determined by the solution of the single particle scattering
problem off a single impurity.  In practice, an appreciable value of
the backscattering amplitude requires the impurity size to be smaller
or of the order of the interparticle spacing, since otherwise the
Fourier component of the potential at $2p_{F}$ is close to zero, and
hence the dimensionless coupling constants $\gamma_{\alpha}\equiv
g_{\alpha}/v_F$ vanish. Therefore, in the following, we will take
$\omega_c \sim \mu$.

The Grassman fields $(\bar{\psi},\psi)$ are free everywhere apart from
the points $x=x_{1,2}$ and hence can be easily integrated out by the
following standard trick: first we formally introduce four
$\delta$-functions into the integrand
\begin{eqnarray}
Z&=& \int \mathrm{D}\bar{\psi} \mathrm{D}\psi \prod_{\alpha=1,2}
\mathrm{D}\bar{\eta}_{\alpha} \mathrm{D}\eta_{\alpha} \,
\delta[\psi(x_{\alpha},\tau)-\eta_{\alpha}(\tau)]\nonumber \\ &\times&
\delta[\bar{\psi}(x_{\alpha},\tau)-\bar{\eta}_{\alpha}(\tau)]
\,e^{-S_\mathrm{FL}-S_{i}},
\end{eqnarray}
where $(\bar{\eta}_{\alpha},\eta_{\alpha})$ are the new Grassman
variables describing the fermions at the location of the individual
impurities. Then we introduce a set of auxiliary fields
$(\bar{\kappa}_{\alpha},\kappa_{\alpha})$ using the identity
$\delta(f)\sim\int \mathrm{D}\kappa\exp(i\int\kappa f\mathrm{d}\tau)$
to rise the $\delta$-functions into the action. Finally, we integrate
out the fermionic fields $(\bar{\psi},\psi)$, which appear only
quadratically, by Fourier transformation:
\begin{equation} 
Z=Z_\mathrm{FL}^0 \int\prod_{\alpha=1,2} \mathrm{D}\bar{\eta}_{\alpha}
\mathrm{D}\eta_{\alpha} \mathrm{D}\bar{\kappa}_{\alpha}
\mathrm{D}\kappa_{\alpha} e^{-S'-S_{i}(\bar{\eta},\eta)},
\end{equation} 
where
\begin{equation}
S_{i}(\bar{\eta},\eta)=\int_{0}^{\beta}\mathrm{d}\tau\sum_{\alpha=1,2}
g_{\alpha}\bar{\eta}_{\alpha}\eta_{\alpha},
\end{equation}
and
\begin{eqnarray}
S'&=&\frac{L}{\beta}\sum_{n}\int\frac{\mathrm{d}p}{2\pi}
\frac{\sum_{\alpha,\beta}\bar{\kappa}_{\alpha}
\kappa_{\beta}e^{ip(x_{\alpha}-x_{\beta})}}
{-i\omega_{n}+\xi_{p}}\nonumber \\ &-&i \sum_{\alpha,n}
(\kappa_{\alpha}\eta_{\alpha} +\bar{\kappa}_{\alpha}
\bar{\eta}_{\alpha}),
\end{eqnarray}
where $\xi_{p}=p^{2}/2m-\mu$, and the summation is over the fermionic
Matsubara frequencies $\omega_{n}$. The trivial prefactor
$Z_\mathrm{FL}^0$ arises from the integration over the fermionic
fields in the absence of impurities, giving the grand partition
function of the homogeneous liquid.  The fields
$(\bar{\kappa}_{\alpha},\kappa_{\alpha})$ depend only on imaginary
time $\tau$, or frequency $\omega_{n}$ in the Fourier representation
and thus the integral over $p$ can be easily calculated.  Since for a
sufficiently large separation $\vert x_1-x_2\vert =r\gg p_{F}^{-1}$,
the interaction energy is small, the relevant frequencies are small
compared to the Fermi energy $\omega_c\sim \mu\sim v_{F} p_F$.  An
expansion to leading order in $\omega_{n}\ll\mu$ then gives
\begin{eqnarray}
S'&=&-\frac{i L}{\beta v_{F}}\sum_{n} \sum_{\alpha,\alpha'} s_n
\bar{\kappa}_{\alpha} \kappa_{\alpha'}
e^{ip_{F}|x_{\alpha}-x_{\alpha'}|s_n} \nonumber \\ &\times&
e^{-|\omega_{n}|(1-\delta_{\alpha,\alpha'})/\omega_{r}}
-i\sum_{\alpha,n} (\kappa_{\alpha}\eta_{\alpha}+ \bar{\kappa}_{\alpha}
\bar{\eta}_{\alpha}),
\end{eqnarray}
where $\omega_{r}\equiv u/r\ll\mu$ and $s_n\equiv
\mathrm{sign}(\omega_n)$.  The characteristic frequency $\omega_{r}$
will play an important role in our subsequent discussions. Physically
it represents the inverse flight time for a characteristic excitation
in the liquid between the locations of the two impurities, which
naturally obeys the inequality $\omega_{r}\ll\omega_{c}$ provided the
impurities are much further apart than the average distance between
two fermions in the liquid.  It is also the quantization energy
between the two impurities.  In order to obtain the impurity
interaction directly from the partition function, we integrate out the
auxiliary fields $(\bar{\kappa}_{\alpha},\kappa_{\alpha})$.  This
results in an action
\begin{equation}
S'(\bar{\eta},\eta)=\frac{\beta v_{F}}{i L} \sum_{n}
  (\bar{\eta}_{1},\bar{\eta}_{2})\left(
\begin{array}{cc}
f_{n} & -f_{n}e_n\\ -f_{n}e_n & f_{n}
\end{array}
\right)\left(
\begin{array}{c}
\eta_{1}\\ \eta_{2}
\end{array}
\right),
\end{equation}
which only depends on the four time dependent Grassman fields
$(\bar{\eta}_{\alpha},\eta_{\alpha})$ which describe the Fermi field
at the impurity positions.  The coefficients $f_{n}$ and $e_{n}$ are
defined by
\begin{equation}
f_{n}\equiv \frac{s_n}{1-e_n^2},\,\, e_n\equiv e^{i s_n p_{F} r -
|\omega_{n}|/\omega_{r}}.
\end{equation}
Including the contribution (\ref{eq:Sint}) due to the interaction
between the impurities and the liquid, the complete expression for the
statistical sum (\ref{eq:statsum}) can now be written as
\begin{equation} 
Z=Z_\mathrm{FL}^0 Z_{\kappa} \int \prod_{\alpha=1,2}
\mathrm{D}\bar{\eta}_{\alpha} \mathrm{D}\eta_{\alpha} e^{-S'-S_{i}},
\end{equation}
where $Z_{\kappa}$ comes from the integration of the auxiliary fields:
\begin{equation}
Z_{\kappa}=\prod_n \left(\frac{i L}{\beta v_{F}} \right)^2 (1-e_n^2).
\end{equation}
The total effective action $S'+S_{i}$ is quadratic in the Grassman
fields $(\bar{\eta}_{\alpha},\eta_{\alpha})$ and thus the integration
can be done exactly to yield $Z=Z_\mathrm{FL}^0 Z_{\kappa} Z_{\eta}$,
with
\begin{equation}
Z_{\eta}=\prod_n \left(\frac{\beta v_{\mathrm{\scriptscriptstyle
        F}}}{i L} \right)^2 \left[ (f_n+i\gamma_1)(f_n+i\gamma_2)-(f_n
        e_n)^2\right]\ .
\end{equation}
Here, the $\gamma_{\alpha}=g_{\alpha}/v_F$ are the dimensionless
backscattering amplitudes, characterizing the interaction of the
impurities and the liquid. We can now obtain the free energy of the 1D
Fermi gas in the presence of the impurities from $F=-\log Z/\beta$ as
follows
\begin{align}
F =& F^0-\frac{1}{\beta} \sum_{n} \log \{(1-e_n^2) \nonumber \\
&\times [(f_n+i\gamma_{1})(f_n+i\gamma_{2})-(f_{n}e_n)^2]\},
\label{F12}
\end{align}
where $F^0=-\log Z_\mathrm{FL}^0/\beta$ is the free energy of the
undisturbed, homogeneous liquid. The expression (\ref{F12}) is ill
defined as it stands, since it contains both the energy of zero-point
fluctuations in the gas, as well as the formally divergent
self-energies of the separate impurities. The relevant interaction
energy associated with a change of the separation of the two
impurities is given by:
\begin{eqnarray}
V_{12} &\equiv& F (\gamma_{\alpha},r)-F(0,r)- [F
(\gamma_{\alpha},\infty)-F (0,\infty)]\nonumber \\ &=&F
(\gamma_{\alpha},r)- F (\gamma_{\alpha},\infty).
\label{renorm}
\end{eqnarray}
The renormalization thus requires subtracting first the free energy of
the liquid without the impurities ($\gamma_{\alpha}=0$, vacuum energy)
and then the free energy of the system when the impurities are very
far apart ($r\rightarrow \infty$, self-energy of the
impurities). While both the vacuum energy and the individual
self-energies are infinite in the absence of a cutoff, the
renormalized interaction (\ref{renorm}) is finite and independent of
the cutoff (see also the discussion below in section \ref{classical}).

At low temperature $T\ll\omega_{r}$ we can switch from summation to
integration according to $\mathrm{d}\omega=2\pi T\mathrm{d}n$, so that
the effective interaction energy between the impurities can be
expressed as:
\begin{equation}
V_{12} =-\int_{0}^{\infty} \frac{\mathrm{d}\omega}{\pi}\log
\left|1+\frac{\gamma_{1}\gamma_{2}e^{-2\omega/\omega_{r}+ 2 i p_{F}r}}
{1+i(\gamma_{1}+ \gamma_{2})- \gamma_{1}\gamma_{2}}\right|.
\label{eq:Vfreefermionsfin}
\end{equation}
The integral can be performed analytically to yield our final result
for the impurity interaction at $T=0$:
\begin{equation}
V_{12}=\frac{v_{F}}{2\pi r}\, \Re\, \mathrm{Li}_2 \left(-\frac{
 \gamma_{1}\gamma_{2} e^{2i p_{F} r}}{1+ i (\gamma_{1} + \gamma_{2})-
 \gamma_{1} \gamma_{2}}\right) \ ,
\label{eq:Dilog}
\end{equation}
where $\mathrm{Li}_2$ is the di-logarithmic function \cite{dilog} and
$\Re$ is the real part.  Obviously the interaction quite generally
falls off very slowly like $1/r$ with an amplitude, which is a
strictly periodic function.  Its period $\pi/p_F=\rho_0^{-1}$ is equal
to the average inter-particle distance. This is a characteristic
property of degenerate fermions, essentially reflecting the well known
Friedel oscillations of the density (see below). Trivially, the
interaction vanishes, if one of the scattering amplitudes
$\gamma_{1,2}$ is zero.

A simple expression for the renormalized interaction energy $V_{12}$
is obtained in two limiting cases. First, if the interaction of the
impurities with the liquid is weak $\gamma_{\alpha}\ll 1$, we can
expand the di-logarithm in Eq.~(\ref{eq:Dilog}) to obtain:
\begin{equation}
V_{12}=-\gamma_{1} \gamma_{2} \frac{v_F}{2\pi r} \cos(2p_{F}r)\, .
\label{eq:V12idealweak}
\end{equation} 
In the limit of strong impurities $\gamma_{\alpha}\gg1$, we find in
turn the result
\begin{equation}
V_{12}=\frac{v_{F}}{2\pi r}\, \Re\, \mathrm{Li}_2 \left(e^{i 2 p_{F}
 r} \right) \ ,
\label{eq:V12limits}
\end{equation} 
which is completely independent of the scattering amplitudes.  In this
case, the interaction energy $V_{12}$ can be represented as
$V_{12}=\frac{v_{\mathrm{\scriptscriptstyle F}}}{2 \pi
r}f(2p_{\mathrm{\scriptscriptstyle F}}r)$, where $f(x)\equiv \Re\,
\mathrm{Li}_2 (e^{i x})$ is a periodic function bounded as follows
$f_{{\min}}\leq f\leq f_{{\rm max}}$ where:
\begin{equation}
f_{{\rm max},{\rm min}}= \mathrm{Li}_2 (\pm 1)
=\frac{\pi^2}{6},-\frac{\pi^2}{12}\ .
\end{equation}
A simple way of understanding the slow $1/r$-decay and the
oscillations with period $\pi/p_F$ may be obtained in the weak
scattering limit Eq.~(\ref{eq:V12idealweak}).  Indeed, the density
perturbation created by a single impurity of strength $\gamma_1$ at
position $x_1$ is asymptotically given by:
\begin{equation} 
\rho_1(x) \approx \rho_0 - \frac{\gamma_1}{2} \frac{\cos(2p_{F}
  |x-x_1|)}{2\pi |x-x_1|}\ .
\end{equation}
This expression for the Friedel oscillations in a spinless 1D
non-interacting Fermi gas is valid in the limit where $\gamma_1 \ll 1$
and $|x-x_1|\gg \rho_0^{-1}$ \cite{Matveev}. Since the impurities
couple to the local density, the interaction energy (excluding
self-energies) of the system of two weak impurities is simply obtained
from $U_{12}(\gamma_{\alpha},r)= g_2\rho_1(x_2)+g_1\rho_2(x_1)$ where
%$r=|x_1-x_2|$ is the distance between the impurities and
$\gamma_{\alpha}\ll 1$. When renormalized
$V_{12}=U_{12}(\gamma_{\alpha},r)- U_{12}(\gamma_{\alpha},\infty)$,
this interaction energy coincides with
Eq.~(\ref{eq:V12idealweak}). Alternatively, the result may be derived
by using the random-phase approximation (RPA) as shown in the Appendix
\ref{app:RPA}.

The analysis in this section is readily generalized to the case of a
Fermi gas with spin. In fact for non-magnetic impurities, such as
considered in the current article, the two spin modes are decoupled
and therefore the energy due to the presence of the two impurities is
simply multiplied by a factor of two.

It should be emphasized that the calculation above, can be immediately
extended to the case of noninteracting fermions in two or three
dimensions $d=2,3$, giving rise to an interaction energy for weak
coupling of the form
\begin{equation}
V_{12}\sim f_{1}(\pi)f_{2}(\pi) \frac{p_F v_F}{(p_{F} r)^d} \cos(2p_{F}r)\, ,
\label{23d}
\end{equation}
where $f_{\alpha}(\pi)$ are the dimensionless backscattering
amplitudes of the impurities. This result follows most simply by
considering the density fluctuations $\delta\rho_{1}(\vec x)$ induced
by a single impurity at position $\vec x_{1}$. As discussed, e.g., in
Ref.~\cite{Zwerger2} they exhibit Friedel oscillations proportional to
the dimensionless backscattering amplitude $f_{1}(\pi)$ at the Fermi
energy. The resulting interaction energy is then simply given by
$V_{12}\propto f_{2}(\pi)\delta\rho_{1}(\vec x_2)$. In fact, this is a
special case of a general result \cite{FerrellLuther}, that the
asymptotic interaction between two impurities is determined by the
product of their backscattering amplitudes. In the presence of short
range repulsive interactions between the fermions, we expect that the
result (\ref{23d}) remains qualitatively correct in the case of two
and three dimensions. This is based on the existence of a Fermi liquid
description in $d=2,3$, which guarantees that the low energy
properties are qualitatively unchanged from those of a Fermi gas. For
example, assuming that the static density response function at $2p_F$
is given by the particle-hole bubble \cite{Negele}, the
renormalization factor $Z<1$ in the single-particle Green function
will give rise to a Fermi liquid correction factor $Z^{2}$ in
$V_{12}$. This argument, however, neglects possible vertex corrections
in the density response which may lead to an enhancement rather than a
suppression of the amplitude of the Friedel oscillations. In fact this
effectively happens in the one-dimensional case, where the vanishing
$Z\,$-factor gives rise to Friedel oscillations, which decay more
slowly than in the noninteracting case (see below). While we are not
aware of a quantitative calculation of the $2p_F\,$-density response
in Fermi-liquids, it is very likely that they will give rise only to
finite, multiplicative corrections to Eq.~(\ref{23d}). As we will see
below, however, the situation in one dimension is quite different from
that in $d=2,3$ in the sense, that even qualitatively the asymptotic
form of the interaction is \emph{not} given by the Friedel oscillation
picture, even for very weak impurities.

Finally, we mention a recent work dealing with neutron matter. Bulgac
\emph{et al.} \cite{Bulgac} consider a neutron star crust, which is
modeled as a degenerate non-interacting neutron gas (i.e. an ideal 3D
Fermi gas) containing various kinds of defects or inhomogeneities
(such as nuclei or bubbles) immersed in it. These authors compute the
interaction energy between two defects resulting from the quantum
fluctuations of the Fermi sea of neutrons. They obtain expressions
similar to Eq.~(\ref{23d}), which can be interpreted as RKKY-like
interactions between defects. In addition, they discuss the influence
of the shape of the defects and consider situations with more than two
defects.

\subsection{Spinless Fermi Luttinger liquid}\label{less}
Realistic Fermi systems consist of interacting particles. In three and
also in two dimensions, it is possible to describe even strong
interactions by Landau's Fermi liquid theory. As is well known,
however, this concept fails in one dimension. Here we consider
fermions with repulsive short-range interaction.  At low-energy such a
system exhibits a gapless excitation spectrum with a linear
dispersion, characteristic for the universality class of Luttinger
liquids (LL) \cite{Haldane1,Haldane2,Giamarchi}.

For simplicity, we start by considering spinless fermions, for which
the low-energy description is given by the following hydrodynamic
action:
\begin{equation}
  S_\mathrm{LL}=\frac{1}{2\pi K}\int \!\!\mathrm{d} x\,\mathrm{d}\tau
  \left[ u
  (\partial_{x}\theta)^{2}+\frac{1}{u}(\partial_{\tau}\theta)^{2}
  \right]. \label{eq:LLaction}
\end{equation} 
Here $u$ is the sound velocity and $K$ the Luttinger parameter. In a
translationally invariant system, they obey the relation $u K= v_F$
\cite{Haldane2}, with $v_F=p_F/m=\pi \rho_0/m$ the Fermi velocity of
the associated non interacting spinless Fermi gas. We consider repulsive 
interactions for which $K<1$. Since the Luttinger liquid description only
applies at low energies, the fields have to be cutoff at energy
$\omega_c \sim \mu$, where $\mu$ is the chemical potential. 
The associated cutoff length $a\equiv u/\omega_c$
is of order $1/\rho_0$. Of course, for a quantitative calculation of
the scale at which the low energy description applies, a microscopic
model is needed, which allows to determine nonuniversal
properties. For single impurities in Luttinger liquids this problem
has only recently been discussed, see \cite{Meden}. Since we are
concerned with the interaction at distances much longer that the
average interparticle separation, only the low energy properties are
relevant, which are well described by the hydrodynamic action
(\ref{eq:LLaction}).  The corresponding field $\theta$ is related to
the density of the liquid by
\begin{equation}
\rho(x)\approx \left(\rho_{0}+ \frac{\partial_x \theta}{\pi}\right)
\left[1+ 2\cos(2\theta+ 2 p_{F} x)\right],
\label{density}
\end{equation}
where $\rho_{0}$ is the equilibrium density and only the first
harmonics are taken into account \cite{Haldane2,KaneFisher}.

The interaction between the impurities and the Luttinger liquid is
taken to be of the form
\begin{equation}
S_{i}=\int_{0}^{\beta}\mathrm{d}\tau\!\sum_{\alpha=1,2}\tilde{g}_{\alpha} 
\rho(x_{\alpha})\, ,
\label{eq:Sidens}
\end{equation}
i.e. a coupling to the local density with phenomenological 
scattering amplitudes $\tilde{g}_{\alpha}$.  Inserting the
expansion (\ref{density}) into this interaction, gives rise to four
different terms. The first term is just the constant Hartree
self-energy of the impurities, which - of course - does not contribute
to the renormalized interaction energy $V_{12}$.  In addition, there
are terms containing $\partial_{x}\theta$ due to forward scattering.
They describe quantum corrections to the self energies but again are
irrelevant for the interaction $V_{12}$ between two widely separated
impurities.  The dominant term for this interaction is the
contribution proportional to $\cos(2\theta+2p_{F}x)$, which is due to
backward scattering. In addition there are higher order terms like
$\partial_{x}\theta \cos(2\theta+2p_{F}x)$, however these are less
relevant in the renormalization group (RG) sense \cite{NoteRG}.
Taking only the most relevant part of the interaction, the coupling
between the impurities and the LL leads to the following nonlinear
contribution to the action
\begin{equation}
S_{i}[\Theta]\approx \int_0^{\beta} \mathrm{d}\tau \sum_{\alpha=1,2} 2
\tilde{g}_{\alpha} \rho_0 \cos[2\sqrt{\pi K}\Theta(x_\alpha)+ 2
p_{F}x_{\alpha}]\ ,
\label{eq:SiTheta}
\end{equation}
where we introduced the renormalized field $\Theta$ by
$\theta=\Theta\sqrt{\pi K}$.  The complete statistical sum of the
system can again be represented by a functional integral. To perform
the integration over the field $\Theta$ we use the same approach as
for the ideal Fermi gas: first we introduce the new variables
$\Theta(x_{1},\tau)=\Theta_{1}(\tau)$ and
$\Theta(x_{2},\tau)=\Theta_{2}(\tau)$ and then insert the two
$\delta$-functions into the functional integral:
\begin{eqnarray}
Z&=& \int \mathrm{D}\Theta \prod_{\alpha=1,2}
\mathrm{D}\Theta_{\alpha} \delta[\Theta(x_{\alpha})-\Theta_{\alpha}]
e^{-S_\mathrm{LL}-S_{i}}.
\end{eqnarray}
We then transform the $\delta$ functions into the functional integrals
over auxiliary fields and perform the Gaussian integration
\begin{equation} 
Z= Z_\mathrm{LL}^0 Z_{\kappa} \int
\mathrm{D}\Theta_{1}\mathrm{D}\Theta_{2}e^{-S_{{\rm
eff}}-S_{i}[\Theta_\alpha]}\label{eq:smallKZ}\ ,
\end{equation}
where the effective action for the real fields $\Theta_{1,2}$ is
\begin{equation} 
S_{{\rm eff}}=\sum_{n} (\Theta_{1,-n},\Theta_{2,-n}) \left(
\begin{array}{cc}
f_{n} & -f_{n}e_n\\ -f_n e_n & f_{n}
\end{array}
\right)\left(
\begin{array}{c}
\Theta_{1,n}\\ \Theta_{2,n}
\end{array}
\right),
\end{equation} 
with the summation occurring over the bosonic Matsubara frequencies
$\omega_{n}$, where $f_{n}\equiv \beta|\omega_{n}|/(1-e_n^2)$ and
$e_n\equiv e^{-|\omega_{n}|/\omega_r}$ with $\omega_r\equiv u/r$.  The
factor $Z_\mathrm{LL}^0$ comes from the integration over
$\Theta(x,\tau)$ and is independent of the impurities, describing the
homogeneous Luttinger liquid. By contrast, the factor
\begin{equation}
Z_{\kappa}=\prod_{n} \left(1-e_n^2\right)^{-1/2}.
\end{equation}
which comes from the integration over the auxiliary fields, describes
the change in the phonon modes due to the constraint on the Fermi
fields at the positions of the two impurities. Similar to the
noninteracting situation, this contribution depends on the associated
characteristic frequency $\omega_{r}$ and is crucial in obtaining a
finite interaction energy $V_{12}$ which is independent of the cutoff.

The remaining and now non-trivial functional integral over the
time-dependent fields $\Theta_{\alpha}$ is of the same form as the one
which appears in the context of quantum Brownian motion in a periodic
potential \cite{Zwerger}. Indeed the effective action (29) basically
describes two quantum particles subject to ohmic dissipation of
dimensionless strength $1/K$ which move in a periodic potential
generated by the backscattering amplitude. As has been shown in
\cite{Zwerger} this problem leads to a localized ground state if
$1/K>1$ with small fluctuations in the field $\Theta$.  For a quantum
liquid with sufficiently strong interactions between the fermions
$K\ll 1$ and strong impurities 
$\tilde{\gamma}_{\alpha}\equiv \tilde{g}_{\alpha}/v_F\gg 1$, the functional
integral (\ref{eq:smallKZ}) over the time-dependent fields
$\Theta_{\alpha}$ can thus be calculated using the stationary phase
approximation (SPA): expanding the functions $\cos(2\sqrt{\pi
K}\Theta_{\alpha}+2p_{F}x_{\alpha})$ from Eq.~(\ref{eq:SiTheta}) to
second order in the fields around one of its minimum, we approximate
the interaction Lagrangian in the form
\begin{equation}
\tilde S_{i}=\int_0^{\beta}
\mathrm{d}\tau\sum_{\alpha=1,2}E_{\alpha}\Theta_{\alpha}^{2}\ ,
\label{eq:statphaseint}
\end{equation}
where $E_{\alpha}=4\pi K \tilde{g}_{\alpha}\rho_0$. Physically this means that
the interaction between each of the impurities and the liquid is
sufficiently strong to pin the local phase $\Theta$ near the value
minimizing the potential energy. The quantities $E_{\alpha}$ play the
role of effective frequencies for the evolution of the fields
$\Theta_{\alpha}$. The approximation of the original Lagrangian
(\ref{eq:SiTheta}) by the quadratic form (\ref{eq:statphaseint}) is
equivalent to an adiabatic approximation which describes physical
processes occurring slower than a time scale given by
$E_{\alpha}^{-1}$. As the typical frequency of interest is $\omega_r$,
the stationary phase approximation is valid when $E_{\alpha}\gg
\omega_r$. It is thus applicable in the case of strong impurities or -
equivalently - long distances (see below) in a strongly repulsive
liquid: $r\rho_0 \tilde{\gamma}_{\alpha}\gg 1/K^2\gg 1$.

Within the quadratic approximation (\ref{eq:statphaseint}), the full
effective action in Eq.~(\ref{eq:smallKZ}) is quadratic in
$\Theta_{\alpha}$ and hence can be evaluated exactly
$Z=Z_\mathrm{LL}^0 Z_{\kappa} Z_{\Theta}$, where:
\begin{equation}
Z_{\Theta}=\prod_{n}\left[(f_n+\beta E_{1}) (f_n+\beta
E_{2})-(f_{n}e_n)^2 \right]^{-1/2}\ .
\end{equation}
The associated free energy $F=-\log Z/\beta $ is given by
\begin{align} 
F=&F^0+\frac{1}{2\beta} \sum_{n} \log \{(1-e_n^2)\nonumber \\ &\times
[(f_n+\beta E_{1})(f_n+\beta E_{2})-(f_{n}e_n)^2 ]\}\ ,
\end{align} 
where $F^0=-\log Z_\mathrm{LL}^0 /\beta$ again describes the
undisturbed, homogeneous liquid.  From the free energy we obtain the
renormalized interaction energy between the two impurities $V_{12}$ in
precisely the same manner as in Eq.~(\ref{renorm}):
\begin{equation}
V_{12}= \frac{1}{\beta} \sum_{n>0} \log\left(1-\frac{E_{1} E_{2}
    e^{-2\omega_n/\omega_{r}}}{\omega_n^{2} + \omega_n
    (E_{1}+E_{2})+E_{1}E_{2}} \right) \ .
\end{equation}
At sufficiently low temperatures $T\ll\omega_{r}$, the summation may
be replaced by an integral over the real frequency $\omega$:
\begin{equation}
V_{12}=\frac{1}{2\pi}\int_0^\infty
\mathrm{d}\omega\log\left(1-\frac{E_{1}E_{2}e^{-2\omega/\omega_{r}}}
{\omega^{2}+\omega(E_{1}+E_{2})+E_{1}E_{2}}\right).
\end{equation}

In the limit of long distances (or strong impurities), i.e.
$E_{\alpha}\gg \omega_{r}$, the integral converges at
$\omega\sim\omega_{r}$ and hence $\omega^{2}\ll\omega E_{\alpha}\ll
E_{\alpha}^{2}$. For sufficiently large separations, we thus obtain
the simple universal interaction
\begin{equation} 
V_{12}=\frac{u}{2\pi r}\int_0^\infty
    \mathrm{d}y\log(1-e^{-2y})=-\frac{\pi}{24}\frac{u}{r}\ ,
\label{eq:V12bosefin}
\end{equation} 
which decays inversely with distance.  The short distance regime,
where $E_{\alpha}\ll \omega_{r}$, can, however, not be considered
within the stationary phase approximation
(\ref{eq:statphaseint}). Indeed, the latter is only justified if the
characteristic energy scale $\omega\lesssim\omega_{r}$ of excitations
involved in the interaction does not exceed the effective frequencies,
i.e. if $E_{\alpha}\gg\omega_{r}$. As a result, for intermediate and
short distances, the interaction energy $V_{12}$ does not follow a
simple $1/r$ -behavior. In particular, it is impossible to describe
the limit $r\to 0$ without properly including a cutoff or working in a
microscopic model from the beginning. It is only within such a more
complete calculation, that the full energy $F
(\tilde{\gamma}_{\alpha},r)-F(0,r)$ of the two impurity problem approaches the
self energy of the doubled single impurity case, as expected on
physical grounds.  For a single scalar field in 1D, as described by
our hydrodynamic action (23), this calculation has recently been done
by Jaffe \cite{Jaffe}.

The validity of the quadratic expansion (\ref{eq:statphaseint}) may be
extended to the whole relevant range $K< 1$ of the Luttinger parameter
with the help of the self-consistent harmonic approximation (SCHA)
\cite{Saito,Zwerger}. In the context of Friedel oscillations around a
single impurity in a Luttinger liquid, this has been used by Egger and
Grabert \cite{Egger}. It is based on making a quadratic approximation
(\ref{eq:statphaseint}) for the backscattering term, however with
frequencies $E_{\alpha}$ which are determined from Feynman's
variational principle.  Using Eq.~(\ref{eq:statphaseint}) as the trial
action, one has to minimize the free energy
\begin{equation}\label{Fvar}
F_\mathrm{var} = -\frac{1}{\beta} \log \tilde{Z} +
\frac{1}{\beta}\langle S - \tilde{S} \rangle_{\tilde{S}} \ ,
\end{equation}
where $S= S_\mathrm{e f f }+S_i$ and $\tilde{S}= S_\mathrm{e f f} +
\tilde{S}_i$. Taking $E_{\alpha}$ as variational parameters, we obtain
\begin{equation} 
\frac{E_{\alpha}}{4\pi K \tilde{g}_{\alpha} \rho_0}=\left( 1+
\frac{\omega_{c}}{E_{\alpha}}\right)^{-K},
\end{equation} 
where $\omega_{c}$ is the high-energy cutoff. Following \cite{Egger},
we define the crossover scale $r_0$ by $E_{\alpha}\equiv u/r_0$ when
both impurities have approximatively the same strength $\tilde{\gamma}_1
\approx \tilde{\gamma}_2 \approx \tilde{\gamma}$. The SCHA is a good approximation
when $K<1$ and $E_{\alpha} \gg \omega_{r}$, i.e.  at long distances
$r> r_0$.

In the limit of strong impurities $\tilde{\gamma}_{\alpha}\gg 1$, the SCHA
frequencies $E_{\alpha}\approx 4\pi K \tilde{g}_{\alpha}\rho_0$ are the same
as those obtained within the stationary phase approximation and the
crossover scale is given by $r_0\rho_0 \sim 1/K^2\tilde{\gamma}$. In the
opposite limit $\tilde{\gamma}_{\alpha}\ll 1$, they are given by
\begin{equation}
E_{\alpha}\approx 4\pi K \tilde{g}_{\alpha}\rho_0\left(\frac{4\pi K
\tilde{g}_{\alpha}\rho_0}{\omega_{c}}\right)^{\frac{K}{1-K}}
\label{crossoverenergy}
\end{equation} 
and the crossover scale is
\begin{equation}
r_0 \rho_0 \sim (K^2\tilde{\gamma})^{-1/(1-K)}(\rho_0 a)^{-K/(1-K)}\ ,
\label{crossoverscale}
\end{equation}
where $\rho_0 a \sim 1$ for repulsive fermions.  Note the singular
behavior of the crossover scale $r_0\to \infty$ in the limit $K\to 1$
of non-interacting fermions.  This implies that the regime of validity
of the SCHA is moved out to extremely long scales $r_0$.  Quite
generally, therefore, the long distance behavior $r\gg r_0$ of the
interaction energy is always given by the Casimir like expression
(\ref{eq:V12bosefin}) whenever $K<1$.  The scale, however, beyond
which this simple result applies, strongly depends on the strength of
the backscattering amplitude and the repulsive interaction.

In order to study the limit of weak impurities and weak interactions
($\tilde{\gamma}_{\alpha}\ll 1$ and $K<1$ close to 1) in more detail, we use
perturbation theory. At second order in $\tilde{\gamma}_{\alpha}$, and when
$r\gg\rho_0^{-1}$, we find
\begin{equation}
V_{12} = - \tilde{\gamma}_1 \tilde{\gamma}_2 (\rho_0 a \pi)^2 
\left(\frac{r}{a}\right)^{2(1-K)} 
\frac{v_F}{2\pi r}\cos (2 p_F r) h(K)\ ,
\label{eq:genFriedel}
\end{equation} 
where $h(K)\equiv \frac{K}{\sqrt{\pi}}\frac{\Gamma(K-1/2)}{\Gamma(K)}$ and 
$\Gamma(z)$ is the Gamma function. The function $h(K)$ diverges as $K\to 1/2$ 
and approaches one at $K= 1$. The above equation was obtained under the assumption 
that $1/2<K<1$. When $K=1$, it reproduces exactly Eq.~(\ref{eq:V12idealweak}) 
obtained for the non-interacting Fermi gas, provided that we choose the 
following relation between the microscopic and 
phenomenological impurity strengths $\gamma_{\alpha}=\tilde{\gamma}_{\alpha} \rho_0 a \pi$.
Perturbation theory breaks down when, in order of magnitude, the
interaction energy (\ref{eq:genFriedel}) reaches the strong impurity
(or long distances) result (\ref{eq:V12bosefin}): $|V_{12}|\sim
\omega_r$. This occurs for $r\rho_0 \sim \tilde{\gamma}^{-1/(1-K)}(\rho_0
a)^{-K/(1-K)}$, in agreement with Eq.~(\ref{crossoverscale}), because
$1/2<K<1$.  Therefore, the perturbative result (\ref{eq:genFriedel})
is valid for intermediate distances $\rho_0^{-1}\ll r \ll r_0$. At
long distances $r\gg r_0$ it is replaced by the Casimir-type
result~(\ref{eq:V12bosefin}).

Finally, it is worth mentioning that the presence of two impurities
implies the possibility of a tunneling resonance \cite{KaneFisher,FN}.
The physics of such a resonance is, however, not captured by the
SCHA. This issue will be discussed in more detail in the Appendix
\ref{app:Coulomb}.

\subsection{Spin-1/2 Fermi Luttinger liquid}\label{spinful}
In this section we generalize the above analysis by including the spin
degree of freedom, again using a Luttinger liquid description.

We consider $N$ fermions in an equal mixture of spin up and spin down
components, i.e., $N_{\uparrow}=N_{\downarrow}=N/2$.  Here the
(non-interacting) Fermi velocity is $v_F=p_F/m=\pi \rho_0 / 2m$ with
$\rho_0 = N/L$. Taking spin into account the Euclidean action of the
Luttinger liquid is generalized to
\begin{equation}
S_\mathrm{LL}[\theta_\mu]=\!\sum_{\mu=\rho,\sigma} \frac{1}{2\pi
  K_\mu}\int \!\!\mathrm{d} x\, \mathrm{d}\tau\, \left[ u_\mu
  (\partial_{x}\theta_\mu)^{2}+\frac{1}{u_\mu}(\partial_{\tau}\theta_\mu)^{2}
  \right]\ ,
\end{equation}
where $[\theta_\mu]$ is an abbreviation for
$[\theta_\rho,\theta_\sigma]$. This corresponds to the usual
``charge'' $\theta_\rho$ and ``spin'' $\theta_\sigma$ fields that are
linear combinations of the spin up and down fields:
$\theta_{\rho/\sigma}= \frac{1}{\sqrt{2}} (\theta_\uparrow \pm
\theta_\downarrow)$. The Luttinger liquid parameters obey the relation
$K_\rho u_\rho = v_{F}$ and in addition
\begin{equation}
K \equiv K_\rho<1 \qquad \textrm{and} \qquad K_\sigma=1,
\end{equation}
where the first equation comes from considering repulsive interactions
and the second from the $SU(2)$ symmetry of the model with
spin-independent interactions \cite{Giamarchi}. In addition $K>1/2$ when 
considering contact interactions between fermions, see e.g. \cite{Recati3}. The action term that
describes the interaction with the impurities is still given by
Eq.~(\ref{eq:Sidens}) where the density is now simply the sum of the
densities of the two spin modes that have the same form as
Eq.~(\ref{density}). As stated in the previous section we only keep
the most relevant term, which corresponds to backscattering by the
impurity, and hence obtain:
\begin{align}
S_i[\theta_\mu] = \sum_{\alpha=1,2} 2 \tilde{g}_{\alpha} \rho_0 \int_0^\beta
\!\!\mathrm{d}\tau \, & \cos [\sqrt{2} \theta_\rho (x_{\alpha}) + 2
p_F x_{\alpha} ] \nonumber \\ & \times \cos [ \sqrt{2} \theta_\sigma
(x_{\alpha}) ] \ .
\end{align}

Now we follow the same procedure as in previous sections and obtain
\begin{equation}
Z= Z^0_\mathrm{LL} Z_{\kappa} \int \! \prod_{\alpha=1,2}
\prod_{\mu=\rho,\sigma}\! \mathrm{D}\Theta_{\mu \alpha}\
e^{-S_\mathrm{e f f}[\Theta_{\mu \alpha} ] - S_i[\Theta_{\mu \alpha}]}
\ ,
\end{equation}
where we have rescaled the fields at the impurity positions by
$\theta_{\mu \alpha}= \sqrt{\pi K_\mu} \Theta_{\mu \alpha}$.  The
effective action $S_{\mathrm{e f f}}$ is given by
\begin{equation}
S_\mathrm{e f f} [\Theta_{\mu \alpha}] = \sum_{n}
\sum_{\mu,\alpha,\delta} I^\mu_{\alpha \delta}\ \Theta_{\mu \alpha,-n}
\, \Theta_{\mu \delta,n} \ ,
\end{equation}
where the $n$ refers to the Matsubara frequencies $\omega_n$ and
$I^\mu_{\alpha \delta}$ are the elements of the matrix
\begin{equation}
I^\mu = f_{\mu n}
\begin{pmatrix}
1 & -e_{\mu n} \\ -e_{\mu n} & 1
\end{pmatrix} \ ,
\end{equation}
with $f_{\mu n}= \beta |\omega_n|/(1-e_{\mu n}^2)$, $e_{\mu n}=
e^{-|\omega_n|/\omega_\mu}$, and $\omega_\mu = u_\mu/r$.

In order to calculate the partition function, we again use the
stationary phase approximation, which corresponds to expanding $S_i$
around its minima to second order in the fields, assuming that the
impurities are strong, i.e.~$\tilde{\gamma}_{\alpha}=\tilde{g}_{\alpha}/v_{F}\gg 1$.
The nonlinear action $S_i$ is thus replaced by a quadratic
approximation
\begin{equation}\label{Sitilde}
\tilde{S}_i= \beta \sum_{\mu \alpha} \sum_n E_{\mu \alpha}
|\Theta_{\mu \alpha,n}|^2 \ ,
\end{equation}
where $E_{\mu \alpha}= 2 \pi K_\mu\,\tilde{g}_{\alpha} \rho_0$.

Since the charge and spin fields are now completely decoupled, we have
$\tilde{Z}= Z_{\rho} Z_{\sigma}$, with:
\begin{equation}
  Z_{\mu} = \prod_n \left[ \left( f_{\mu n } +\beta E_{\mu 1} \right)
      \left( f_{\mu n}+ \beta E_{\mu 2} \right) -(f_{\mu n} e_{\mu
      n})^2 \right]^{-1/2} \ .
\end{equation}
The total free energy $F=F_\rho+F_\sigma$ is then simply the sum of
the charge and spin contribution.

After renormalization the free energy is given by:
\begin{equation}
V_{12} = \frac{1}{\beta} \sum_{\mu} \sum_{n>0} \log \left[1-
\frac{E_{\mu 1} E_{\mu 2} \ e^{-2 \omega_n/\omega_\mu}}{(\omega_n^2 +
E_{\mu 1}) (\omega_n^2 + E_{\mu 2})} \right] \ .
\end{equation}
At low temperature $T\ll\omega_\mu$ we can again replace the sum over
Matsubara frequencies by an integral. In the limit of strong
impurities $\omega_\mu\ll E_{\mu a}$ we obtain
\begin{equation}
V_{12} = \sum_{\mu} u_\mu \int_0^\infty \! \frac{\mathrm{d} y}{2 \pi}
\log (1- e^{-2 y}) = -\frac{\pi}{24}\frac{u_\rho+u_\sigma}{r}.
\label{eq:CasSpin}
\end{equation}
This is the straightforward generalization for spin $1/2$ fermions of
the result obtained in the previous section, see
Eq.~(\ref{eq:V12bosefin}). As discussed there, the SPA is valid only
if $\omega_\mu\ll E_{\mu a}$.  Hence we are unable to calculate the
interaction at shorter distances, where $\omega_\mu\gg E_{\mu a}$.

As in the case of spinless fermions, we can go beyond the SPA regime,
using the SCHA. In particular we use $S= S_\mathrm{e f f }+S_i$ and
$\tilde{S}= S_\mathrm{e f f} + \tilde{S}_i$ in Eq.~(\ref{Fvar}); and
then we minimize $F_\mathrm{var}$ with respect to $E_{\mu a}$. In the
case of identical impurities, i.e.~$\tilde{\gamma}_1=\tilde{\gamma}_2=\tilde{\gamma}$, the
values of $E_{\mu \alpha}$ that minimize $F_\mathrm{var}$ are such
that $E_{\mu 1} = E_{\mu 2} = E_{\mu}=\pi K_\mu E$. For large
distances, i.e., for $\omega_\mu \ll E_\mu$, the SCHA is valid and $E$
is given by
\begin{equation}
E = 2 \tilde{g} \rho_0 \left( 1+ \frac{\omega_c}{\pi K E}\right) ^{-K/2}
\left( 1+ \frac{\omega_c}{\pi K E}\right)^{-1/2} \ ,
\end{equation}
where $\omega_c$ is the high energy cutoff.  As in the spinless case,
we can define the crossover scale as $r_0=\max(u_\mu/E_\mu)=v_{F}/\pi
K^2 E$, since $K<1$.  In the limit of very strong impurity
backscattering $E_\mu\gg \omega_c$, we recover the SPA result, i.e.
$E_\mu=2 \pi K_\mu \tilde{g} \rho_0$, and the crossover scale is $r_0
\rho_0\sim 1/K^2 \tilde{\gamma}$. For intermediate impurity strength
$\omega_\mu \ll E_\mu \ll \omega_c$, we obtain
\begin{equation}
E_\rho = E_\sigma/K = 2 \pi \tilde{g} \rho_0 K^\frac{1}{1-K} \left(\frac{2 \pi
\tilde{g} \rho_0}{\omega_c} \right)^{\frac{1+K}{1-K}} \ ,
\end{equation}
and the crossover scale in this case is given by
\begin{equation} \label{spincross}
\rho_0 r_0 \sim \tilde{\gamma}^{-\frac{2}{1-K}} K^{-\frac{3}{1-K}}
(a\rho_0)^{-\frac{1+K}{1-K}} \ ,
\end{equation}
where $a\equiv v_{F} /K\omega_c$ is the short distance cutoff and $a\rho_0
\sim 1$. Note that, as in the spinless case, the crossover scale
diverges $r_0\to \infty$ when $K\to 1$.  For weak interactions,
therefore, the result (\ref{eq:CasSpin}) is only valid at very
large distances.

The regime of weak impurity strength can be studied using perturbation
theory. At second order in $\tilde{\gamma}_{\alpha}$, we find
\begin{equation}\label{spinpert}
V_{12} = - \tilde{\gamma}_1 \tilde{\gamma}_2 (\rho_0 a \pi)^2 \left(\frac{r}{a}\right)^{1-K} 
\frac{v_F}{\pi r}\cos (2 p_F r) h(K)\ ,
\end{equation}
where $h(K)\equiv K^K
\frac{\Gamma(K/2)}{\sqrt{\pi}\Gamma(\frac{K+1}{2})} {}_2F_{1} \left(
\frac{K}{2},\frac{K}{2}; \frac{K+1}{2};1-K^2 \right)$ and ${}_2F_{1}$
is the hyper\-geometric function \cite{hypergeo}. The function $h(K)$
is a smooth, monotonically decreasing function of $K$ which diverges
like $h(K)\sim 2/\pi K$ as $K\to 0$ and approaches one at $K= 1$. 
When $K=1$, the preceding result reproduces exactly Eq.~(\ref{eq:V12idealweak}) 
(with an extra factor of two due to the spin degeneracy) obtained for the non-interacting
(spinless) Fermi gas, provided that we choose the following relation between the microscopic and 
phenomenological impurity coupling constants $\gamma_{\alpha}=\tilde{\gamma}_{\alpha} \rho_0 a \pi$. 
The perturbative result (\ref{spinpert})
is valid for $r \ll r_0$, with $r_0$ given in
Eq.~(\ref{spincross}). At the crossover scale $r=r_0$ and for
$K\lesssim 1$, $|V_{12}|\sim \omega_r$ as in
Eq.~(\ref{eq:CasSpin}). In conclusion, therefore, the interaction
between two impurities follows the behavior given by
Eq.~(\ref{spinpert}) only for intermediate distances and weak
interactions.  By contrast, for large distances, there is a crossover
to the universal Casimir-type interaction Eq.~(\ref{eq:CasSpin}) which
depends only on the velocities $u_{\rho}$ and $u_{\sigma}$.

\subsection{Discussion}\label{sec:discussion}
As our main result, we have shown that for separations much larger
than the interparticle spacing, the interaction energy of two
impurities in a Luttinger liquid of repulsive fermions ($K<1$) is a
Casimir-type interaction, given by a very simple universal relation,
Eq.~(\ref{eq:V12bosefin}) (resp. Eq.~(\ref{eq:CasSpin})) for spinless
(resp. spin $1/2$) fermions.  In contrast to the result
Eq.~(\ref{eq:V12limits}) obtained for strong impurities in a
non-interacting Fermi gas it does not contain Friedel oscillations and
is independent of both the impurity strengths and the interaction
parameter $K$.  The physical origin of this long range force is thus
quite different from the $K=1$ case. In the non-interacting gas, the
long range force comes from the polarization of the ground state. In
the strongly interacting case, the Friedel oscillations of the ground
state density around each of the independent impurity still exist
\cite{Egger} but they are not relevant for the impurity interaction at
long distances.  Instead the result Eq.~(\ref{eq:V12bosefin}) is best
understood as being the Casimir interaction energy of two mirrors,
i.e. impenetrable impurities, in a phononic bath \cite{Casimir}. This
interpretation is supported by the direct calculation of the
interaction energy of two mirrors in the vacuum fluctuations of a 1D
scalar field which represents the density modes of the intervening
quantum liquid, see, e.g., Ref.~\cite{Zee}. In fact, a similar result
has previously been found for the force between two infinite mass
beads on a string \cite{DHokerSikivie}.  The Friedel oscillations are
relevant for the interaction between two impurities only in the
non-interacting case or at intermediate distances in the interacting
Luttinger liquid ($K<1$), see Eq.~(\ref{eq:genFriedel}) and
(\ref{spinpert}).

The resulting picture is consistent with the RG calculation of Kane
and Fisher for a single impurity \cite{KaneFisher}: when $K<1$ and
$\tilde{\gamma}>0$, the backscattering amplitude is renormalized to strong
coupling in the low energy (or long distance $r\gg r_0$) limit.  The
liquid is thus effectively cut into pieces, with the impurities acting
like perfect mirrors for the acoustic modes, resulting in a Casimir
force between them. The scale on which the impurities flow to strong
coupling depends on: i) the initial strength of the impurities
$\tilde{\gamma}$ and ii) the flow velocity given by $1-K$, see below. When the
impurities are strong and the liquid is strongly interacting, the
impurities flow quickly to strong coupling.  The associated crossover
scale $r_0$ is thus of the order of the interparticle distance. By
contrast, when the impurities are weak and the liquid is almost
non-interacting, it takes very long distances to reach the asymptotic
regime. Qualitatively, the crossover scale $r_0$ for two weak
impurities can already be obtained from the scaling theory for a
single impurity. Indeed, the perturbative flow equation of Kane and
Fisher \cite{KaneFisher} gives as the running impurity strength:
\begin{equation} 
\tilde{\gamma}_{\mathrm{eff}}\approx \tilde{\gamma} (r/a)^{1-K}\ .
\end{equation}
For spin $1/2$ fermions, the preceding equation holds provided that 
the exponent $1-K$ is replaced by $(1-K)/2$. 
The crossover scale $r_0$ then corresponds to the distance at which
the running impurity strength $\tilde{\gamma}_{\mathrm{eff}}$ becomes of order
one, i.e. $r_0 \rho_0 \sim \tilde{\gamma}^{-1/(1-K)}$, in agreement with
Eq.~(\ref{crossoverscale}), because $\rho_0 a \sim 1$ and $1/2<K<1$
for repulsive fermions with contact interactions. For longer
distances, the impurity reaches strong coupling and cuts the liquid
into disconnected pieces.

\section{One-dimensional Bose liquid}
In this section, we discuss the case of two identical impurities $g\delta(x_{\alpha})$ in a
one-dimensional Bose liquid. In particular, we consider 1D bosons with
short range repulsive interactions $g_B \delta(x)$. As was first shown
by Lieb and Liniger \cite{LiebLiniger}, the dimensionless interaction
parameter $\gamma_B\equiv mg_B/\rho_0$ in this problem is inversely
proportional to the density. The strong coupling, Tonks-Girardeau,
limit $\gamma_B\gg 1$ is thus reached either for strong repulsion or
at low densities. Within a low-energy effective Luttinger liquid
description, the Luttinger parameter $K$ for interacting bosons is
larger than $1$. It is related to the sound velocity $u$ by $u=\pi
\rho_0/mK $. In the weakly interacting, Gross-Pitaevskii limit
$\gamma_B \ll 1$, the Luttinger parameter diverges like
$K\approx\pi/\sqrt{\gamma_B}\to \infty$. For strong coupling $\gamma_B
\gg 1$ in turn, one finds $K\approx 1+4/\gamma_B \to 1$. The singular
case of non-interacting bosons ($\gamma_B=0$) is discussed separately
in Appendix \ref{app:nibosons}.

\subsection{Classical ground state energy and vacuum fluctuations}\label{classical}
Before starting the explicit calculation of the renormalized impurity
interaction energy $V_{12}(r)$ in the case of bosons, we discuss a
limitation of our quantum hydrodynamic approach when treating the Bose
case. We will shortly see, why this limitation was not discussed in
the context of fermions. In quantum hydrodynamics, the ground state
energy $E_0$ is obtained as the sum of two terms: the classical ground
state energy $E_0^{cl.}$ and the quantum vacuum fluctuations
$E_0^{q.fl.}$, see e.g. Ref.~\cite{AGD}. The classical ground state
energy $E_0^{cl.}[\rho_0(x)]$ is a functional of the density profile
$\rho_0(x)$.  Now, the presence of impurities or boundaries in a
quantum liquid modifies the density profile $\rho_0(x)$ over distances
of the order of the healing length $\xi\equiv 1/\sqrt{m\mu}$. This in
turns modifies the classical ground state energy in addition to the
change in the quantum vacuum fluctuations, which are responsible for
the Casimir-type interaction energy. In the LL approach, the classical
ground state energy is usually neglected and only the fluctuations
\emph{above} the classical ground state are taken into account. In a
homogeneous system, the classical ground state energy is just a
constant in the Hamiltonian and can therefore be safely
ignored. Provided that the healing length is much smaller than the
system size, the effect of the boundaries disappears for bulk
properties. In fermionic quantum liquids, the healing length is of the
order of the interparticle distance $1/\rho_0$. Therefore, introducing
two impurities with a separation $r\gg 1/\rho_0$ leaves the classical
ground state energy unchanged. However, in the case of bosons, the
healing length is much larger than the interparticle spacing,
diverging like $\xi\sim K/\rho_0$ for weak interactions $\gamma_B\to
0$. The condition $r\gg\xi$ for neglecting the contribution of the
classical ground state energy in the calculation of the renormalized
interaction energy between impurities $V_{12}$ thus becomes
increasingly restrictive for small $\gamma_B$. As we will see, this
leaves us only with an exponentially small interaction, quite in
contrast to the case of fermions.

\subsection{Weak impurities and weakly interacting bosons}
In the case of a weakly interacting Bose gas $\gamma_B\ll 1$
(corresponding to \emph{high} densities) the Bogoliubov approach is
quantitatively applicable, as was shown long ago by Lieb and Liniger
\cite{LiebLiniger}. The physical reason for that is, that the healing
length $\xi\approx 1/\rho_0\sqrt{\gamma_B}$ in this limit is much
larger than the average interparticle spacing. The situation is
therefore essentially equivalent to the weak coupling limit of a 3D
Bose-Einstein condensate at \emph{low} densities, where interactions
between two impurities have recently been discussed by Klein and
Fleischhauer \cite{KleinFleishhauer} (the interaction energy $V_{12}$
is called the conditional energy shift and denoted by $\Delta$ in this
work). The explicit calculation is based on the Bogoliubov approach
and assumes that the dimensionless impurity strength $\gamma\equiv
mg/\pi\rho_0$ is much smaller than one. Adapting the calculation of
Ref.~\cite{KleinFleishhauer} to a one-dimensional gas, we obtain
\begin{equation}
V_{12}(\gamma,r)\approx -\gamma^2\rho_0 \xi \frac{\pi^2
\rho_0^2}{m}\exp{(-2r/\xi)}
\end{equation}
to lowest order in $\gamma$. The interaction energy between widely
separated impurities in a weakly interacting Bose gas thus vanishes
exponentially on the scale set by the healing length $\xi$. As
discussed above, this calculation assumes that the impurities modify
the density profile only locally, which is also the physical reason
for the vanishing interaction at $r\gg\xi$.
 
The preceding result can again be qualitatively understood with the
help of the scaling theory of Kane and Fisher \cite{KaneFisher} for a
single weak impurity in a LL. Indeed, when $K>1$, the effective
coupling of a single impurity is renormalized to zero in the
low-energy limit. The RG flow thus starts at high-energy
$\omega_c=u/a\sim \mu$ (corresponding to a short distance cutoff
$a\sim \xi$) and ends at a much lower energy $\omega_r=u/r$ which is
set by the separation $r$ of the two impurities. Similar to the case
of fermions, we define a crossover scale $r_0$ by the condition that
the effective dimensionless impurity strength at this scale is of
order one:
\begin{equation}
\gamma_{\mathrm{eff}}\approx \gamma (a/r)^{K-1}\sim 1 \Leftrightarrow
r_0\approx \xi \gamma^{1/(K-1)} \lesssim \xi \ .
\end{equation}
Since $\gamma \ll 1$ for weak impurities and $K=\pi/\sqrt{\gamma_B}\to
\infty$ due to the weak interaction condition, we find $r_0 \approx
\xi$. This confirms that there is no interaction between two weak
impurities embedded in a weakly interacting Bose gas when they are
further apart than a distance of the order of the healing length. More
generally, the scaling theory indicates that there is no long range
interaction between weak impurities even in a strongly interacting
Bose gas, in which $K\gtrsim 1$. Of course the limiting case $K=1$ of
hard-core bosons is special as is the case $K=\infty$ of no
interaction at all. The latter case is treated in Appendix
\ref{app:nibosons} and shows, that there is no interaction between the
impurities whatever the impurity strength. In strong contrast to that,
the limit of a Tonks-Girardeau gas of hard-core bosons is equivalent
to the case of non-interacting fermions for properties depending only
on the modulus of the ground state wavefunction like the density
distribution. On the basis of the calculations in section
\ref{nonint}, one thus expects a long-range interaction of the form
(\ref{eq:Dilog}) between impurities in a Tonks-Girardeau gas which
exhibits Friedel-like oscillations. In view of the fact that the
momentum distribution of hard-core bosons is quite different from that
of a free Fermi gas, showing no jump at $p_F$, this is a quite
remarkable result \cite{Caza}. Another singular limit, where
long-range interactions appear in a Bose liquid is that of
impenetrable impurities $\gamma = \infty$. For arbitrary values
$\infty> K>1$ of the interactions, the interacting Bose liquid is then
cut in three disconnected pieces. The impurities thus act as perfect
mirrors for the low energy phonon excitations of the intervening Bose
liquid, giving rise to a Casimir interaction energy precisely as in
Eq.~(\ref{eq:V12bosefin}) for spinless fermions. It appears, however,
that the limit of impenetrable impurities at arbitrary energies is
nonphysical, imposing strict Dirichlet boundary conditions on a scalar
field \cite{Jaffe}. The Casimir force is thus expected to be
restricted to $\gamma=\infty$, while for any finite $\gamma$ only
short range interactions should survive. Describing these crossovers
in detail, clearly requires a quantitative theory of impurity
interactions in 1D Bose liquids at arbitrary values of $K$ and
$\gamma$, an interesting problem for further study.

\section{Experimental realization and detection of the Casimir-like force}
The recent realization of one-dimensional ultracold Fermi gases in a
strong 2D optical lattice \cite{Esslinger} provides a novel
opportunity to study Luttinger liquid effects in a setup with cold
gases, e.g. spin-charge separation \cite{Recati2}. In order to study
whether the Casimir interactions discussed here might be observed in
these systems, we consider an atomic gas of fermions in two hyperfine
states. These two internal states play the role of (iso)spin 1/2
states. In principle both the sign and the strength of the interaction
can be controlled using scattering resonances, e.g. a confinement
induced resonance as shown by Olshanii \cite{Olshanii}. For
simplicity, we assume that the two spin states are equally populated
$N_{\uparrow}=N_{\downarrow}=N/2$. We note that in order to have $K<1$, 
one needs to consider a two-component atomic Fermi gas. Indeed, in a single 
component Fermi gas, Pauli's principle forbids $s$-wave collisions implying 
that $K=1$ for spinless fermions interacting via a contact potential.

Following several recent ideas \cite{Recati,KleinFleishhauer,Raizen}
which involve trapping single atoms in ultracold gases, we consider an
atomic quantum dot (AQD) like configuration, which consists of single
atoms confined in a tight trap created either magnetically or
optically, e.g. by an additional optical lattice. We assume that the
confining potential can be adjusted in such a way, that it does not
affect the atoms of the bath.  The impurity atom, which is trapped in
a certain internal state $|a\rangle$ interacts with the atoms of the
bath through $s$-wave collisions.  In the case where two such AQDs are
embedded in the bath and both impurity atoms are in state $|a\rangle$,
the system precisely realizes the situation of two localized
impurities interacting via a 1D quantum liquid. Provided the liquid
consists of spin-1/2 repulsive fermions, we expect that for distance
$r$ much larger than the average interparticle spacing, the
interaction is of the Casimir form given in Eq.~(\ref{eq:CasSpin}).
In principle, using a scattering resonance may allow to reach the
strong impurity regime $\tilde{\gamma}\gg 1$ where the crossover scale $r_0$
is even smaller than the inter-particle distance $1/\rho_0$.
\begin{figure}[ptb]
\includegraphics[height=6cm]{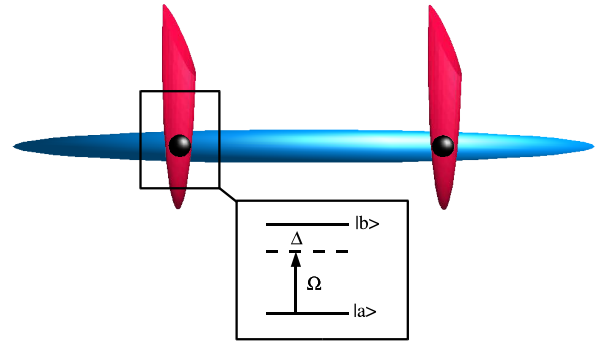}
\caption{Schematic setup of 2 AQDs coupled to a 1D atomic reservoir.
The impurity atoms (see text) in tightly confining potential interact
with the bath when their internal level is $|a>$. Here $\delta$ is the
renormalized detuning and $\Omega$ is the Rabi frequency coming from a
laser induced coupling, see section \ref{conclusion}.}
\label{scheme}
\end{figure}

A possible way to detect the interaction energy $V_{12}(r)$, is to do
spectroscopy of a single trapped atom as a function of the distance
$r$ to a neighboring trapped atom. In addition to the mean field line
shifts modifying the internal levels of the impurity atom, the Casimir
interaction produces a line shift depending on distance as $1/r$. For
a quantitative estimate of this effect, we compute the energy
(\ref{eq:CasSpin}) for the experimental situation realized in
Ref. \cite{Esslinger}. There, about $N\sim 100$ $^{40}$K atoms (per
tube) form an atomic wire of length $L\sim 10$~$\mu$m. The temperature
can be as low as $T\sim 50$~nK, which is about one tenth of the Fermi
temperature. The Fermi velocity is of order $v_F \sim 2.10^{-2}$~m/s
and we take $u_{\sigma}+u_{\rho} \sim 2v_F$. As the tube length is of
the order of 10~$\mu$m and the inter-particle distance $1/\rho_0 \sim
0.1$~$\mu$m, we assume the inter-impurity distance to be $r \sim
1$~$\mu$m, which is larger than the crossover length for strong
impurities. This gives a Casimir-related line shift of the order of
$1$~kHz, which is in an experimentally accessible range. With the
parameters given above, the characteristic frequency $\omega_r$ is of
the order of $\sim 3T$, which is not much larger than $T$ as required
for the validity of the zero temperature limit in which the Casimir
force is obtained.

\section{Conclusion}\label{conclusion}
In conclusion, we studied the long-range interaction between two
impurities mediated by the 1D quantum liquid in which they are
embedded. We found that for repulsive fermions, the impurities
interact via a RKKY-like interaction at intermediate distances and via
a Casimir-like force at large distances. The crossover scale
separating these two regimes depends on the strength of the impurities
and on the interactions between fermions. We proposed an experimental
realization of such a system with atomic quantum dots in an ultra-cold
atomic gas and suggested a way to detect the Casimir-type interaction
by spectroscopy of a single atom in an AQD.

An issue which is still open is to understand the interaction between
impurities in a strongly interacting Bose liquid. In particular, how
does the short-range interaction (on the scale of the healing length)
turn into a long-range interactions featuring Friedel oscillations in
the Tonks-Girardeau limit? Another issue is to assess the validity of
the self-consistent harmonic approximation (SCHA) used to discuss spin
$1/2$ fermions in section \ref{spinful}. Indeed, it is not obvious
that the variational ansatz, which assumes decoupling of the charge
and spin modes in presence of the impurities, is a valid starting
point.

In this paper, we studied static impurities. The situation becomes
even more interesting if one has dynamic impurities, as in an
AQD. First, we discuss the possibility of \emph{internal} dynamics for
the AQD. We have seen that the characteristic frequencies of vacuum
modes (excitations) responsible for the long-range interactions
between AQDs are limited by $\omega_{r}$. This means that the
effective interaction potential between static impurities can be used
(in an adiabatic approximation) for time-dependent impurity strengths
$\tilde{\gamma}_{\alpha}(t)$ provided that the interaction properties change
slowly compared with the time scale $\omega_{r}^{-1}$. Consider, for
instance, the configuration with two AQDs described previously, see
Figure \ref{scheme} (a similar scheme for two impurities in a 3D
Bose-Einstein condensate is discussed in
Ref.~\cite{KleinFleishhauer}). The two-level impurity atoms can be
described as isospins $1/2$ or qubits. A laser can drive transitions
(equivalent to single-qubit gate) between the two internal levels
$|a\rangle$ and $|b\rangle$. In the adiabatic approximation, the AQD
variables are slow and can be taken out of the integrals so that our
previous treatment to calculate the interaction between two impurities
applies. Thus one can easily write an effective Hamiltonian for the 2
AQDs in the form
\begin{eqnarray}
H_{{\rm
eff}}&=&\sum_{\alpha=1,2}\left(-\frac{\delta}{2}\sigma_{z}^{(\alpha)}+\Omega\sigma_{x}^{(\alpha)}\right)
\nonumber\\
&+&\frac{1}{2}V_{12}(\sigma_{z}^{(1)}+1)(\sigma_{z}^{(2)}+1)\ ,
\label{eq:Hent}
\end{eqnarray}
where $\delta$ is the (renormalized) detuning, $\Omega$ is the
(effective) Rabi frequency coming from the laser induced coupling
\cite{Recati} and $\sigma_x^{(\alpha)}$, $\sigma_y^{(\alpha)}$,
$\sigma_z^{(\alpha)}$ are the Pauli matrices describing the isospin
$1/2$ of each AQD $\alpha=1,2$. The long-range potential $V_{12}$
depends, as we have seen, strongly on the characteristics of the bath
and on the distance between the AQDs. In addition, the case of
impurities with internal dynamics embedded in a spin $1/2$ fermionic
liquid is of course also relevant for discussing the RKKY interaction
in Luttinger liquids, as discussed perturbatively in
\cite{EggerSchoeller}.

It is also possible to imagine \emph{external} motion or dynamics for
the impurities. The argument of adiabaticity also holds in the case
where the distance between the impurities changes sufficiently
slowly. This means that the Casimir-like interaction could be used,
for example, to create long range attractive forces in mixtures of 1D
fermionic gases. However, when the external dynamics of the impurities
are taken into account on the same footing as the bath dynamics other
effects could reduce, if not wash out, the Casimir force
\cite{CastroNeto}.

\section*{Acknowledgments}
This work was started in collaboration with Piotr Fedichev, whose
substantial contributions to the ideas presented here are gratefully
acknowledged. It is a pleasure also to thank Prof. Alan Luther for
helpful remarks on his unpublished work and Mauro Antezza, Iacopo
Carusotto, Peter Kopietz, Walter Metzner, In\`es Safi and Peter Zoller
for useful discussions. Laboratoire de Physique des Solides is a mixed
research unit (UMR 8502) of the CNRS and the Universit\'e Paris-Sud
$11$ in Orsay. A.R. has been also supported by the European Commission
through contract IST-2001-38863 (ACQP).

\appendix

\section{RPA for weak impurities in an ideal Fermi gas} \label{app:RPA}
In this appendix, we give another derivation of the weak coupling
result (\ref{eq:V12idealweak}) using the random phase approximation
(RPA). The interaction between the two impurities may be written as
the sum of two contributions: a direct interaction (which is zero in
the present case due to the short range nature of the impurities), and
an indirect interaction induced via the polarization of the
medium. The polarization operator in the Fourier representation
$\Pi(\omega_{n},q)$ has a singularity at $q=2p_F$, which leads to the
appearance of the long range force. Specifically, we calculate the
interaction energy of the two impurities using the RPA, which works
well whenever only states with very few particle-hole pairs are
excited by the perturbation, i.e. in the limit $\gamma_{\alpha}\ll
1$. For an ideal (spinless) Fermi gas the RPA Lagrangian can be
represented as ($T=0$ and here we use the real time formalism from the
start):
\begin{equation}
S^{{\mathrm{RPA}}}=\frac{1}{2}\int
\mathrm{d}t\sum_{k}\sum_{p}(\dot{\phi}_{pk}^{\dagger}\dot{\phi}_{pk}-
\omega_{pk}^{2}\phi_{pk}^{\dagger}\phi_{pk}),
\end{equation}
where $\rho_{k}=\sum_{p}(2\omega_{pk})^{1/2}\phi_{pk}$ is the RPA
expression for the fermion density,
$\omega_{pk}=(p+k)^{2}/2m-p^{2}/2m$ is the energy of the electron-hole
pair, and the summation over the momentum $p$ is limited by the
conditions: $|p|<p_{F}$ and $|p+k|>p_{F}$. The interaction of the AQDs
with the liquid can be written as
\begin{equation}
S_{i}^{{\mathrm{RPA}}}=\frac{1}{2}\sum_{pk}(2\omega_{pk})^{1/2}(\phi_{pk}
V_{k}+\phi_{pk}^{\dagger}V_{-k}),
\end{equation} 
where $V_{k}=g_{1}+g_{2}\exp(ikr)$. The total RPA action
$S^{{\mathrm{RPA}}}+S_{i}^{{\mathrm{RPA}}}$ is quadratic, and hence
the fields $\phi_{pk}$ can be integrated out, so that the effective
interaction between the impurities is given by
$V_{12}=-\frac{1}{2}\sum_{k}|V_{k}|^{2}\Pi(\omega=0,k),$ where the
polarization operator is:
\begin{equation}
  \Pi(0,k)=-\frac{2}{\pi}\int_{-p_{\mathrm{\scriptscriptstyle
        F}}}^{p_{\mathrm{\scriptscriptstyle
        F}}}\mathrm{d}p\frac{1}{2pk+k^{2}}=-\frac{1}{\pi
        k}\log\left|\frac{k+2p_{\mathrm{\scriptscriptstyle
        F}}}{k-2p_{\mathrm{\scriptscriptstyle F}}}\right|\ .
\end{equation} 
Substituting $V_{k}$, performing the integration by transforming the
integral along the real $k$ axis into the integrals along the
branch cut corresponding to the singularity of the $\log$-function,
and removing the self energies of the separated impurities
(renormalizing the interaction), we find
\begin{equation}
V_{12}=-\gamma_{1}\gamma_{2} \frac{v_F}{2 \pi r} \cos(2 p_{F}r)\, ,
\end{equation} 
in agreement with Eq.~(\ref{eq:V12idealweak}).

\section{Coulomb blockade and resonant tunneling}\label{app:Coulomb}
The present paper discusses the interaction energy between two
impurities in an atomic quantum wire (i.e., a 1D quantum liquid made
of atoms). In this appendix, we wish to make contact with related
subjects in the field of mesoscopic conductors (or solid-state quantum
wires): namely Coulomb blockade and resonant tunneling of electrons in
a quantum wire with two tunnel junctions or barriers (see, e.g.,
\cite{KaneFisher,FN,Enss}). For simplicity, we only consider the case of
spinless fermions.

We first consider an atomic quantum wire with two impurities.  When
the impurities are strong (or very far apart), the atomic quantum wire
is cut in three disconnected pieces and we may picture the low energy
behavior of the system as the following structure: left
wire/impurity/island(or central wire)/impurity/right wire.  The
Casimir effect occurring in this system is directly related to the
energy cost of transferring a supplementary particle from one of the
wires to the island. In our case (neutral atoms interacting via a
short-range potential), this energy cost is a finite size energy
\cite{Haldane-finitesize} equal to:
\begin{equation}
\pi v_F/2K^2r\ .
\label{finitesizeenergy}
\end{equation}
With the help of the approximate relation $K^{-2}\approx 1+g_{f}/\pi
v_F$ \cite{KaneFisher}, this energy cost can be seen as the sum of two
contributions: a kinetic energy cost $\pi v_F/2r$ and an interaction
energy cost, resulting from the local interaction with the other
particles on the island, $g_{f}/2r$, where $g_{f}$ is the forward
scattering coupling constant (in the standard notation of the g-ology,
$g_{f}=g_2=g_4$, see \cite{Giamarchi}, e.g.). Note that the finite
size energy is of the order of the zero-point kinetic energy of a
phonon on the island $\pi v_F/2Kr$.

In the case of electrons in a solid-state quantum wire with two tunnel
junctions (corresponding to the structure: left
electrode/barrier/island/barrier/right electrode), the Coulomb
blockade is due to the energy cost of transferring a single electron
from an electrode to the island \cite{KaneFisher,FN}.  However, this
energy cost is not only due to the finite size energy $\pi v_F/2K^2r$
(with $K^{-2}\approx 1+e^2/\epsilon \pi v_F$, where $e$ is the
electron charge and $\epsilon$ is an appropriate dielectric constant)
but also gets an additional contribution from the charging energy
$e^2/2C$ of the capacitors (i.e. the two tunnel junctions), where $C$
is the sum of the capacitance of each tunnel junction \cite{FN}. The
total energy cost is equal to $\pi v_F/2K^2r+e^2/2C$.  The charging
energy is due to the long-range part of the interaction between
electrons and therefore does not arise in the case of cold atoms
interacting via a short-range potential.

Another subject of comparison between the atomic quantum wire and the
solid-state quantum wire is the possibility of tunneling resonances
across the double barrier structure \cite{KaneFisher,FN} (see
\cite{Giamarchi} for review).  Indeed, in an atomic quantum wire with
two impurities, when $\cos(p_F r)=0$ \cite{KaneFisher,FN} a tunneling
resonance occur: the energy cost to add a particle on the island
vanishes and particles can therefore tunnel trough the impurities. The
liquid is no more cut into disconnected pieces and we do not expect a
Casimir effect to occur.  Tunneling resonances do not appear in our
calculations because they are not captured by the self-consistent
harmonic approximation we used, as discussed in Ref.~\cite{Egger} for
example. However such resonances are infinitely sharp at zero
temperature \cite{KaneFisher,FN} and therefore they do not play a
major role and should be easy to avoid experimentally.

\section{Impurities in an ideal Bose gas}\label{app:nibosons}
The ideal Bose gas ($\gamma_B=0$) is a singular case: the healing
length $\xi$ diverges and boundaries are therefore felt over
macroscopic distances. Here using quantum hydrodynamics makes no
sense, but of course one can exactly solve the problem from first
principles (Schr\"odinger equation). At zero temperature, all bosons
are in the same single-particle wavefunction $\psi_0$ (the
corresponding many-body wavefunction is just a product or Hartree
state), which is the ground state of the single-particle Schr\"odinger
equation
\begin{equation}
-\frac{1}{2m}\partial_x^2\psi_0(x)
+g[\delta(x-r/2)+\delta(x+r/2)]\psi_0(x)= \epsilon_0\psi_0(x),
\end{equation}
where $g$ is the impurity coupling constant. 
We assume that the particles are on a ring of length $L$. The ground
state energy $E_0$ for $N$ bosons is given by $N\epsilon_0$. For
$g\geq 0$, it is bounded as follows
\begin{equation}
\epsilon_0(g=0,r)\leq \epsilon_0(g,r)\leq \epsilon_0(g=\infty,r),
\end{equation}
which is a direct consequence of the Schr\"odinger equation.  In the
following we will show that both bounds are going to zero in the
thermodynamic limit, implying that $E_0(g,r)=0$ for all $g\geq
0$. Therefore, the interaction energy $V_{12}(g,r)\equiv
E_0(g,r)-E_0(g,r\to \infty)$ vanishes whatever the distance between
the impurities.

On the one hand, when the impurity strength is zero $g=0$, the ground
state of the single-particle Schr\"odinger equation with periodic
boundary conditions is just the constant wavefunction, which has zero
energy $\epsilon_0(r,g=0)=0$. On the other hand, when $g=\infty$, the
wavefunction has to vanish on the location of the impurities, implying
some bending of the wavefunction and a corresponding cost in kinetic
energy. The ground state wavefunction is
\begin{eqnarray}
\psi_0(x)&=&\psi_m\sin \left(\frac{\pi(|x|-r/2)}{L-r}\right) \text{
if } |x|>r/2\nonumber \\ &=& 0 \text{ if } |x|\leq r/2
\end{eqnarray}
where $\psi_m\equiv \psi_0(x=\pm L/2)$ is the maximum value of the
wavefunction and the energy is:
\begin{equation}
\epsilon_0(g=\infty,r)=\frac{1}{2m}\left(\frac{\pi}{L-r}\right)^2
\end{equation}
This quantity vanishes in the thermodynamic limit ($L\to \infty$ at
fixed density $\rho_0=N/L$) such that $L\gg r$. Therefore
$\epsilon_0(g=\infty,r)=0$ in the thermodynamic limit, for all $r$
such that $r\ll L$.  Of course, this conclusion does not hold for an
ideal Fermi gas, because fermions have to occupy different
single-particle states (following the Pauli principle) and therefore
the average energy per particle does not vanish in the thermodynamic
limit.

In conclusion, there is no interaction energy between two impurities
in an ideal Bose gas, provided that the distance $r$ is much smaller
than the ring size $L$, which is always satisfied in the thermodynamic
limit.


\begin{thebibliography}{99}

\bibitem{Moritz} H. Moritz, T. St\"oferle, M. K\"ohl, and
T. Esslinger, Phys. Rev. Lett. \textbf{91}, 250402 (2003).

\bibitem{Paredes} B. Paredes, A. Widera, V. Murg, O. Mandel,
S. F\"olling, I. Cirac, G. V. Shlyapnikov, T. W. H\"ansch, and
I. Bloch, Nature {\bf 429}, 277 (2004).

\bibitem{Weiss} T. Kinoshita, T. Wenger, and D. S. Weiss, Science {\bf
305}, 1125 (2004).

\bibitem{Esslinger} H. Moritz, T. St\"oferle, K. G\"unter, M. K\"ohl,
and T. Esslinger, Phys. Rev. Lett. \textbf{94}, 210401 (2005).

\bibitem{Recati} A. Recati, P. O. Fedichev, W. Zwerger, J. von Delft,
and P. Zoller, Phys. Rev. Lett. \textbf{94}, 040404 (2005).

\bibitem{Marcus}M. J. Biercuk, S. Garaj, N. Mason, J. M. Chow, and
C. M. Marcus, Nano Lett. \textbf{5}, 1267 (2005).

\bibitem{Haldane1}F. D. M. Haldane, J. Phys. C \textbf{14}, 2585 (1981).

\bibitem{Haldane2}F. D. M. Haldane, Phys. Rev. Lett. \textbf{47}, 1840
(1981).

\bibitem{KaneFisher}C. L. Kane, and M. P. A. Fisher, Phys. Rev.
Lett. \textbf{68}, 1220 (1992); Phys. Rev. B \textbf{46}, 15233
(1992).

\bibitem{Enss}T. Enss, V. Meden, S. Andergassen, X. Barnab\'e-Th\'eriault, 
W. Metzner, and K. Sch\"onhammer, Phys. Rev. B \textbf{71}, 155401 (2005).

\bibitem{Popov}V. N. Popov, \emph{Functional integrals in quantum
field theory and statistical physics} (Reidel, Dordrecht, Holland,
1983).

\bibitem{Girardeau}M. D. Girardeau, J. Math. Phys. \emph{1}, 516
(1960); M. D. Girardeau, Phys. Rev. \emph{139}, B500 (1965).

\bibitem{dilog}M. Abramowitz, and I. A. Stegun eds., \emph{Handbook of
Mathematical Functions with Formulas, Graphs, and Mathematical Tables}
(Dover, New York, 1972), \S 27.7; see also
http://mathworld.wolfram.com/Dilogarithm.html

\bibitem{Matveev}K. A. Matveev, D. Yue, and L. I. Glazman,
Phys. Rev. Lett. \textbf{71}, 3351 (1993).

\bibitem{Zwerger2} W. Zwerger, L. B\"onig, and K. Sch\"onhammer,
Phys. Rev. B \textbf{43}, 6434 (1991).

\bibitem{FerrellLuther} R. A. Ferrell, and A. Luther, unpublished
(1967).  The argument is based on calculating the energy-momentum
tensor half way between the impurities, where it is determined by the
asymptotic form of the scattering states. We are grateful to
Professor A. Luther for a discussion on this point.

\bibitem{Negele}J. W. Negele, and H. Orland, \emph{Quantum
Many-Particle Systems} (Addison Wesley, Redwood City, CA, 1988).

\bibitem{Bulgac}A.~Bulgac, and P.~Magierski, Nucl. Phys. A{\bf 383},
695 (2001); A.~Bulgac, and A.~Wirzba, Phys. Rev. Lett. {\bf 87},
120404 (2001).

\bibitem{Giamarchi}T. Giamarchi, \emph{Quantum Physics in One
Dimension} (Oxford University Press, Oxford, 2004).

\bibitem{Meden}V. Meden, W. Metzner, U. Schollw\"ock, and
K. Sch\"onhammer, Phys. Rev. B \textbf{65}, 045318 (2002).

\bibitem{NoteRG}This is true when the impurities are weak and a
perturbative RG applies, but is hard to justify in the opposite
limit. We also neglect higher harmonics, which again relies on the
perturbative approach based on a bosonized theory, as is used here,
see \cite{KaneFisher}.

\bibitem{Zwerger}M. P. A. Fisher, and W. Zwerger, Phys. Rev. B,
\textbf{32}, 6190 (1985).

\bibitem{Jaffe}R. L. Jaffe, AIP Conf. Proc. \textbf{687}, 3 (2003); 
e-print hep-th/0307014.

\bibitem{Saito}Y. Saito, Z. Phys. B \textbf{32}, 75 (1978).

\bibitem{Egger}R. Egger, and H. Grabert, Phys. Rev. Lett, \textbf{75},
  3505 (1995); in \emph{Quantum Transport in
    Semiconductor Submicron Structures, NATO Advanced Study Institute,
    Series E: Applied Sciences, Vol. 326}, edited by B. Kramer
  (Kluwer, Dordrecht, 1996).

\bibitem{FN}A. Furusaki, and N. Nagaosa, Phys. Rev. B, \textbf{47},
3827 (1993).

\bibitem{Recati3}A. Recati, P. O. Fedichev, W. Zwerger, and P. Zoller,
J. Opt. B: Quantum Semiclass. Opt. \textbf{5}, S55 (2003).

\bibitem{hypergeo}M. Abramowitz, and I. A. Stegun eds., \emph{Handbook
of Mathematical Functions with Formulas, Graphs, and Mathematical
Tables} (Dover, New York, 1972), \S 15; see also
http://mathworld.wolfram.com/HypergeometricFunction.html

\bibitem{Casimir}H. B. G. Casimir, Proc. K. Ned. Akad. Wet. \textbf{51},
793 (1948).

\bibitem{Zee}A. Zee, \emph{Quantum Field Theory in a Nutshell}, pages
65-67 (Princeton University Press, Princeton, 2004).

\bibitem{DHokerSikivie}E.~D'Hoker, and P.~Sikivie,
Phys. Rev. Lett. \textbf{71}, 1136 (1993).

\bibitem{LiebLiniger}E. H. Lieb, and W. Liniger, Phys. Rev.
\textbf{130}, 1605 (1963).

\bibitem{AGD}A. A. Abrikosov, L. P. Gorkov, and I. E. Dzyaloshinski,
\emph{Methods of Quantum Field Theory in Statistical Physics} (Dover,
New York, 1975), \S 1.2.
  
\bibitem{KleinFleishhauer} A. Klein, and M. Fleischhauer, Phys. Rev. A
\textbf{71}, 33605 (2005).

\bibitem{Caza}M. A. Cazalilla, J. Phys. B:
At. Mol. Opt. Phys. \textbf{37}, S1-S47 (2004); 
Europhys. Lett. \textbf{59}, 793 (2002).

\bibitem{Recati2}A. Recati, P. O. Fedichev, W. Zwerger, and P. Zoller,
Phys. Rev. Lett.  \textbf{90}, 020401 (2003).

\bibitem{Olshanii}M. Olshanii, Phys. Rev. Lett. \textbf{81}, 938
(1998).

\bibitem{Raizen}B. Mohring, M. Bienert, F. Haug, G. Morigi,
W. P. Schleich, and M. G. Raizen, Phys. Rev. A \textbf{71}, 053601 
(2005).

\bibitem{EggerSchoeller}R.~Egger, and H.~Schoeller, Phys. Rev. B
\textbf{54}, 16337 (1996).

\bibitem{CastroNeto}A. H. Castro Neto, and M. P. A. Fisher,
Phys. Rev. B \textbf{53}, 9713 (1996).

\bibitem{Haldane-finitesize} Such a finite size energy was first
discussed by Haldane \cite{Haldane2} in a slightly different
context. He noticed that for a finite size Luttinger liquid, phonons
are not the only type of low-energy excitations, but that there are
also two so-called ``zero modes''. One of the zero modes corresponds
to adding a particle to the system without adding current. For a
system of size $r$, the corresponding energy cost is $\pi v_N/2r$
(where $v_N=u/K=v_F/K^2$ is a characteristic velocity depending on the
compressibility of the system, which was introduced in
Ref.~\cite{Haldane2}). This is precisely the energy cost of
transferring a particle from one of the wires to the island, see
Eq.~(\ref{finitesizeenergy}).


\end{thebibliography}
\end{document}